\documentclass[aps,twocolumn,pra,showpacs,amsmath,amssymb,showkeys,10pt]{revtex4-2}
\usepackage{graphicx}
\usepackage{epstopdf}
\pdfoutput=1
\usepackage{braket}
\usepackage{hyperref}
\usepackage{longtable}
\usepackage{diagbox}

\begin{document}

	\title{Investigating entropic dynamics of multiqubit cavity QED system}

	\author{Hui-hui Miao}
	\email[Correspondence to: Vorobyovy Gory 1, Moscow, 119991, Russia. E-mail address: ]{hhmiao@cs.msu.ru}
	\affiliation{Faculty of Computational Mathematics and Cybernetics, Lomonosov Moscow State University, Vorobyovy Gory 1, Moscow, 119991, Russia}

	\date{\today}

	\begin{abstract}	
	Entropic dynamics of a multiqubit cavity quantum electrodynamics system is simulated and various aspects of entropy are explored. In the modified version of the Tavis--Cummings--Hubbard model, atoms are held in optical cavities through optical tweezers and can jump between different cavities through the tunneling effect. The interaction of atom with the cavity results in different electronic transitions and the creation and annihilation of corresponding types of photon. Electron spin and the Pauli exclusion principle are considered. Formation and break of covalent bond and creation and annihilation of phonon are also introduced into the model. The system is bipartite. The effect of all kinds of interactions on entropy is studied. And the von Neumann entropy of different subsystems is compared. The results show that the entropic dynamics can be controlled by selectively choosing system parameters, and the entropy values of different subsystems satisfy certain inequality relationships.
	\end{abstract}


	\maketitle

	\section{Introduction}
	\label{sec:Introduction}
	
	The idea of entropy was initially introduced to thermodynamics in 1865 by Clausius \cite{Brush1976}. Subsequently, Boltzmann established the relationship between entropy and the system's microscopic characteristics and gave it statistical significance. Entropy is a measure of the degree of disorder in a microscopic system used in statistical physics. In 1929, Szilard provided an explanation of the connection between information and entropy. Entropy is used as a measure of information in information theory. Later, von Neumann expanded classical entropy to the quantum domain \cite{VonNeumann1932}. And entropy is used to investigate entanglement in the interaction between light and matter \cite{Phoenix1988, Phoenix1990, Phoenix1991, Buzek1992, Buzek1993, Knight1993}. Shannon then developed information theory based on Boltzmann's classical entropy \cite{Shannon1948}. On the other hand, the phenomena of quantum entanglement has garnered a lot of attention with the advancement of quantum mechanics and the founding of quantum information science. Einstein and other scientists were the first to observe the phenomena of quantum entanglement \cite{Einstein1935, Schrodinger1935}. Quantum entanglement is currently widely applied in quantum communications and quantum information processing. Measuring the quantum entanglement between two-body pure state systems is one of the key functions of quantum entropy. To this day, the quantum entropy can be found using a variety of quantifiers, including information entropy \cite{AboKahla2015, AboKahla2018, AboKahlaFarouk2019}, conditional entropy \cite{Bennett1995, Cerf1997}, relative entropy \cite{Vedral1998}, compressed entropy \cite{Fang2000, FangZhou2000}, differential entropy \cite{Yao2018}, etc \cite{Costa2002, Bandt2002, Keller2009, Obada2009, AboKahla2012, AbdelAty2013, Bandt2016, AboKahla2016, AboKahla2019}. In recent years, the wide application of quantum entropy theory in the fields of quantum optics and quantum information has attracted great attention. It is the theoretical basis and powerful tool for understanding hot topics of quantum mechanics and quantum informatics \cite{Sete2015, Bogolyubov2020, Zidan2021, ZidanAbdelAty2021, AboKahla2021}.
	
	A key contribution of this research is the model of quantum electrodynamics (QED), which offers a unique scientific framework for studying the interaction between matter and light. Fields of cavities in this paradigm are coupled to impurity two- or multi-level structures, usually called atoms. The Jaynes--Cummings model (JCM) \cite{Jaynes1963} and Tavis--Cummings model (TCM) \cite{Tavis1968} and their generalizations --- Jaynes--Cummings--Hubbard model (JCHM) and Tavis--Cummings--Hubbard model (TCHM) \cite{Angelakis2007}, which are usually easier to achieve in experiment, are the most researched cavity QED models. Recent years have seen a significant amount of study in the field of cavity QED models \cite{WeiHuanhuan2021, Prasad2018, GuoLijuan2019, Smith2021, Ozhigov2019, Dull2021, Ozhigov2021, Miao2023, MiaoOzhigov2023, MiaoHuihui2024, Li2024}.

	Although concept of entropy is useful in classical physics, it is challenging to extend it to quantum mechanics. Thus, people have tried to investigate various aspects of quantum entropy in QED systems in recent years \cite{Zidan2022, Chandran2019, Najera2020, Fedida2024, Ji2019}. Nevertheless, most prior research has primarily focused on entropic dynamics in simple QED systems, with a rare exception of more complicated systems. Recently, one of the current advancements in quantum technology is to use quantum information technology to simulate chemical and even biological reactions \cite{Zhu2020, Wang2021, McClean2021, Ozhigov2021, Claudino2022, Miao2023}, involving multiqubit systems. Studying the entropic dynamics of these systems will facilitate the future research on more complex reactions. Consequently, it is imperative to study entropic dynamics in multiqubit systems.
	
	This paper is organized as follows. After introducing the target model --- the neutral hydrogen molecule association-dissociation model with covalent bond and phonon in Sec. \ref{sec:Model}, we introduce numerical method in Sec. \ref{subsec:PTSIM} and definitions of von Neumann entropy in Sec. \ref{subsec:Entropy}. We present the effect of interactions on entropy and comparison of entropy of different subsystems in Secs. \ref{subsec:Effect} and \ref{subsec:Comparison}, respectively. Some brief comments on our results and extension to future work in Sec. \ref{sec:ConcluFuture} close out the paper. Some technical details are included in Appendices \ref{appx:RWA}, \ref{appx:Operators} and \ref{appx:Reduced}. 
	
	\begin{figure*}
		\centering
		\includegraphics[width=1.\textwidth]{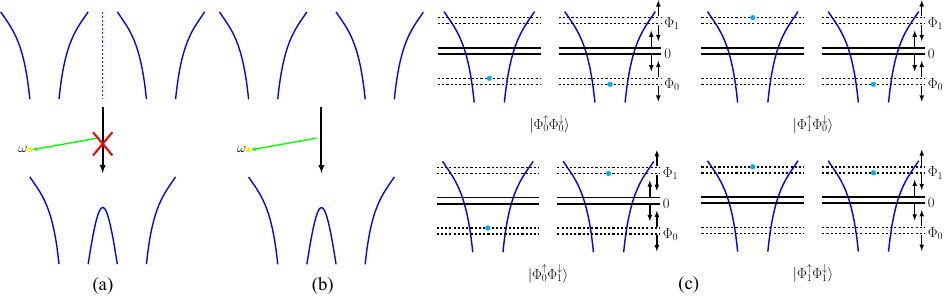}
		\caption{(online color) {\it The model with covalent bond and phonon.} When atoms are in distinct cavities depicted in panel (a), hybridization of atomic orbitals is prohibited; however, when atoms are in the same cavity depicted in panel (b), hybridization of atomic orbitals is allowed. As atomic orbitals hybridize, a covalent bond is formed and a phonon is released; conversely, when a covalent bond is broken, a phonon is absorbed. In panel (c), initial state of entire system, denoted by $|\Psi_{initial}\rangle$, is $|0_1^{\uparrow}\rangle_e|0_2^{\downarrow}\rangle_e$. According to Eqs. \eqref{eq:MolState} and \eqref{eq:AtState}, $|\Psi_{initial}\rangle$ can be decomposed into the sum of four states $|\Phi_0^{\uparrow}\Phi_0^{\downarrow}\rangle$, $|\Phi_1^{\uparrow}\Phi_0^{\downarrow}\rangle$, $|\Phi_0^{\uparrow}\Phi_1^{\downarrow}\rangle$ and $|\Phi_1^{\uparrow}\Phi_1^{\downarrow}\rangle$ (see Eqs. \eqref{eq:InitialDecomposed} and \eqref{eq:PhiPhi}). In the panels, the blue and yellow dots, respectively, stand for electrons and photons. The vertical red dashed line in panel (a) indicates a significant distance between the cavities.}
		\label{fig:Model}
	\end{figure*}
	
	\section{The neutral hydrogen molecule association-dissociation model with covalent bond and phonon} 
	\label{sec:Model}
	
	The association-dissociation model of the neutral hydrogen molecule, studied in detail in our previous work on quantum evolution \cite{Miao2023, MiaoOzhigov2023}, is modified from the TCHM. In this paper, we directly quote the model called the neutral hydrogen molecule association-dissociation model with covalent bond and phonon from \cite{Miao2023} and study the entropy dynamics and its various aspects in the non-dissipative case, that is to say, in a closed quantum system. So in this paper we do not use the quantum master equation (QME) --- Lindbladian.
	
	There are two hydrogen atoms in the model, which form hydrogen bond through orbital hybridization and release a phonon. On the contrary, when the hydrogen bond is broken, phonon need to be absorbed. Both the association and the dissociation reactions can be interpreted by this target model. Each energy level in the molecular--atomic system --- both atomic and molecular --- is divided into two levels, spin up and spin down, denoted by the signs $\uparrow$ and $\downarrow$, respectively. There will now be twice as many levels, and the Pauli exclusion principle \cite{Pauli1925} dictates that each level can only contain one electron. Electron spin flips are temporarily ignored. The excited states of the electron with the spins for the first nucleus are indicated by $|0_1^{\uparrow}\rangle_e$ and $|0_1^{\downarrow}\rangle_e$. Usually simply written as $|0_1\rangle_e$, which can denote both $|0_1^{\uparrow}\rangle_e$ and $|0_1^{\downarrow}\rangle_e$. For the second nucleus --- $|0_2\rangle_e$, which can denote both $|0_2^{\uparrow}\rangle_e$ and $|0_2^{\downarrow}\rangle_e$. Hybridization is impossible when two atoms are in different cavities (see Fig. \ref{fig:Model} (a)). Only when two atoms are in the same cavity, orbital hybridization is possible (see Fig. \ref{fig:Model} (b)).
	
	Hybrid molecular states of the electron are denoted by 
	\begin{subequations}
		\label{eq:MolState}
		\begin{align}
			&|\Phi_1\rangle_e=\frac{1}{\sqrt{2}}\left(|0_1\rangle_e-|0_2\rangle_e\right)\label{eq:MolStatePhi1}\\
			&|\Phi_0\rangle_e=\frac{1}{\sqrt{2}}\left(|0_1\rangle_e+|0_2\rangle_e\right)\label{eq:MolStatePhi0}
		\end{align}
	\end{subequations}
	where $|\Phi_1\rangle_e$ is molecular excited state, $|\Phi_0\rangle_e$ is molecular ground state. On the contrary, we can get the atomic state after orbital de-hybridization
	\begin{subequations}
		\label{eq:AtState}
		\begin{align}
			&|0_1\rangle_e=\frac{1}{\sqrt{2}}\left(|\Phi_0\rangle_e+|\Phi_1\rangle_e\right)\label{eq:AtState1}\\
			&|0_2\rangle_e=\frac{1}{\sqrt{2}}\left(|\Phi_0\rangle_e-|\Phi_1\rangle_e\right)\label{eq:AtState2}
		\end{align}
	\end{subequations}

	The Hilbert space of quantum states of the entire system, having the following form
	\begin{equation}
		\label{eq:SpaceBondPhonon}
		|\Psi\rangle_{\mathcal{C}}=|p_1\rangle_{\Omega^{\uparrow}}|p_2\rangle_{\Omega^{\downarrow}}|m\rangle_{\omega}|l_1\rangle_{\Phi_1^{\uparrow}}|l_2\rangle_{\Phi_1^{\downarrow}}|L\rangle_{cb}|k\rangle_{n}
	\end{equation}
	where $p_1,\ p_2$ are the numbers of molecular photons with modes $\Omega^{\uparrow}$, $\Omega^{\downarrow}$, respectively; $m$ is the number of phonons with mode $\omega$. $l_1,\ l_2$ describe orbital state: $l_1=1$ --- electron with spin $\uparrow$ in excited orbital $\Phi_1^{\uparrow}$, $l_1=0$ --- electron with spin $\uparrow$ in ground orbital $\Phi_0^{\uparrow}$; $l_2=1$ --- electron with spin $\downarrow$ in excited orbital $\Phi_1^{\downarrow}$, $l_2=0$ --- electron with spin $\downarrow$ in ground orbital $\Phi_0^{\downarrow}$. The states of the covalent bond are denoted by $|L\rangle_{cb}$: $L=0$ --- covalent bond formation, $L=1$ --- covalent bond breaking. The states of the nuclei are denoted by $|k\rangle_n$: $k=0$ --- state of nuclei, gathering together in one cavity, $k=1$ --- state of nuclei, scattering in different cavities.
	
	Hamiltonian of this new model with consideration of rotation wave approximation (RWA, see Appx. \ref{appx:RWA}) has following form
	\begin{equation}
		\label{eq:HamilBondPhonon}
		\begin{aligned}
			H&=\hbar\Omega^{\uparrow}a_{\Omega^{\uparrow}}^{\dag}a_{\Omega^{\uparrow}}+\hbar\Omega^{\downarrow}a_{\Omega^{\downarrow}}^{\dag}a_{\Omega^{\downarrow}}+\hbar\omega a_{\omega}^{\dag}a_{\omega}\\
			&+\hbar\Omega^{\uparrow}\sigma_{\Omega^{\uparrow}}^{\dag}\sigma_{\Omega^{\uparrow}}+\hbar\Omega^{\downarrow}\sigma_{\Omega^{\downarrow}}^{\dag}\sigma_{\Omega^{\downarrow}}+\hbar\omega\sigma_{\omega}^{\dag}\sigma_{\omega}\\
			&+g_{\Omega^{\uparrow}}\left(a_{\Omega^{\uparrow}}^{\dag}\sigma_{\Omega^{\uparrow}}+a_{\Omega^{\uparrow}}\sigma_{\Omega^{\uparrow}}^{\dag}\right)\sigma_{\omega}\sigma_{\omega}^{\dag}\\
			&+g_{\Omega^{\downarrow}}\left(a_{\Omega^{\downarrow}}^{\dag}\sigma_{\Omega^{\downarrow}}+a_{\Omega^{\downarrow}}\sigma_{\Omega^{\downarrow}}^{\dag}\right)\sigma_{\omega}\sigma_{\omega}^{\dag}\\
			&+g_{\omega}\left(a_{\omega}^{\dag}\sigma_{\omega}+a_{\omega}\sigma_{\omega}^{\dag}\right)\\
			&+\zeta\left(\sigma_n^{\dag}\sigma_n+\sigma_n\sigma_n^{\dag}\right)
		\end{aligned}
	\end{equation}
	where $\hbar=h/2\pi$ is the reduced Planck constant. $g$ is the coupling strength between the photon or phonon (with annihilation and creation operators $a$ and $a^{\dag}$, respectively) and the electron in the molecule (with excitation and relaxation operators $\sigma^{\dag}$ and $\sigma$, respectively). $g_{\Omega^{\uparrow}}$ is the coupling strength for the photon mode $\Omega^{\uparrow}$, $g_{\Omega^{\downarrow}}$ is the coupling strength for the photon mode $\Omega^{\downarrow}$, $g_{\omega}$ is the strength of formation or breaking of covalent bond, $\zeta$ is the tunneling strength. $\sigma_{\omega}\sigma_{\omega}^{\dag}$ verifies that covalent bond is formed. The matrix form of all operators are shown in Appx. \ref{appx:Operators}.
	
\begin{figure}
		\centering
		\includegraphics[width=0.4\textwidth]{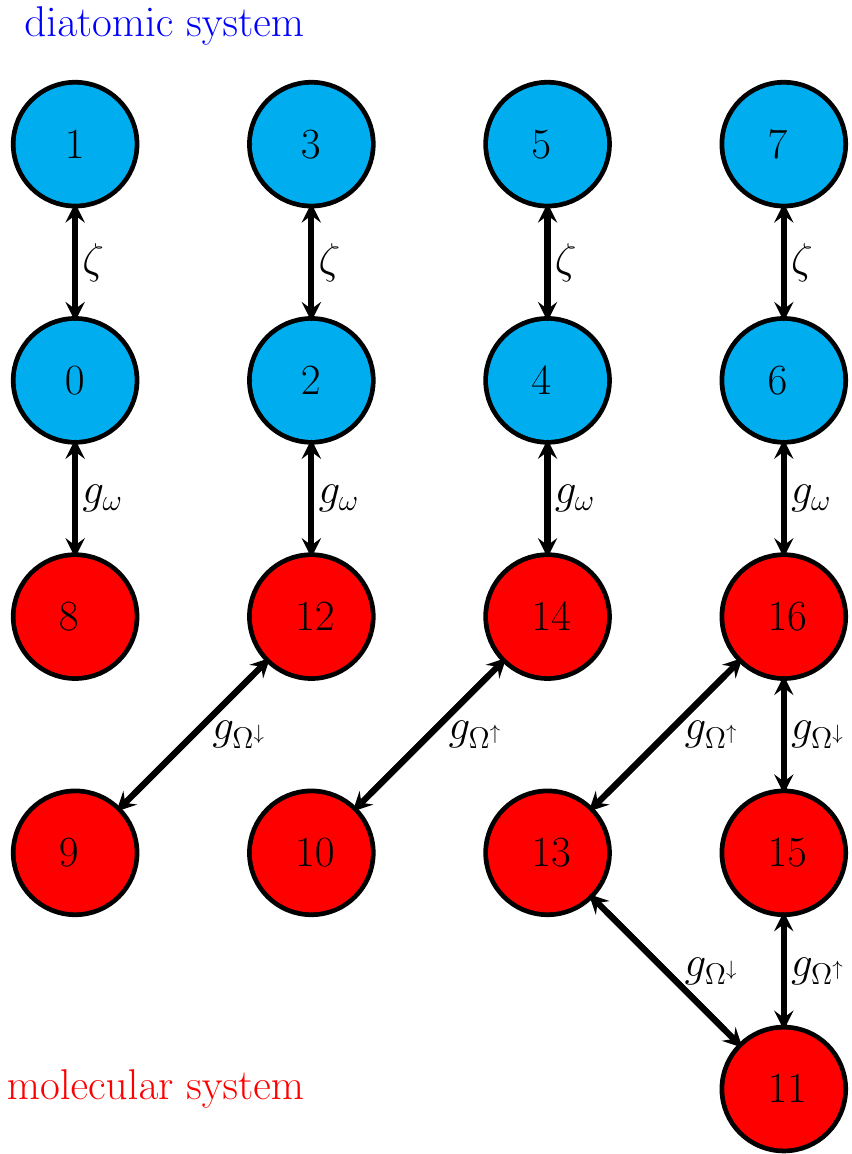}
		\caption{(online color) {\it Interactions between quantum states.} In this figure, four types of interactions are shown: $g_{\Omega^{\uparrow}}$, $g_{\Omega^{\downarrow}}$, $g_{\omega}$ and $\zeta$. The blue circles correspond to the break of the covalent bond, that is, $|L\rangle_{cb}=|1\rangle_{cb}$, which is a diatomic system. The red circles correspond to the formation of the covalent bond, that is, $|L\rangle_{cb}=|0\rangle_{cb}$, which is a molecular system. Initially, system state consists of four states $|0\rangle$, $|2\rangle$, $|4\rangle$ and $|6\rangle$, corresponding to $|\Phi_0^{\uparrow}\Phi_0^{\downarrow}\rangle$, $|\Phi_0^{\uparrow}\Phi_1^{\downarrow}\rangle$, $|\Phi_1^{\uparrow}\Phi_0^{\downarrow}\rangle$ and $|\Phi_1^{\uparrow}\Phi_1^{\downarrow}\rangle$, respectively, which are defined in Eq. \eqref{eq:PhiPhi}.}
		\label{fig:Interaction}
	\end{figure}

	For this seven-qubit system of $2^7$ states, in fact, only a small part of the states participate in the evolution, because we do not consider the dissipative process. In addition, the initial state and some conditions also limit the scale of the states participating in the evolution. We can generate these states through two steps:
	\begin{itemize}
		\item generating and numbering potential evolution states involved in the evolution in accordance with the initial state;
		\item establishing Hamiltonian with these states and potential interactions among them.
	\end{itemize}
	Using this technique, 17 states are obtained, which are shown in Tab. \ref{tab:States} and interactions between these states are shown in Fig. \ref{fig:Interaction}.
	\begin{table}[!htpb]
        \centering
		\begin{tabular}{|c|c|}
			\hline
			\diagbox{Index}{State}{Hilbert space} & $|p_1\rangle|p_2\rangle|m\rangle|l_1\rangle|l_2\rangle|L\rangle|k\rangle$ \\
			\hline
			0 & $|0\rangle|0\rangle|0\rangle|0\rangle|0\rangle|1\rangle|0\rangle$ \\
			\hline
			1 & $|0\rangle|0\rangle|0\rangle|0\rangle|0\rangle|1\rangle|1\rangle$ \\
			\hline
			2 & $|0\rangle|0\rangle|0\rangle|0\rangle|1\rangle|1\rangle|0\rangle$ \\
			\hline
			3 & $|0\rangle|0\rangle|0\rangle|0\rangle|1\rangle|1\rangle|1\rangle$ \\
			\hline
			4 & $|0\rangle|0\rangle|0\rangle|1\rangle|0\rangle|1\rangle|0\rangle$ \\
			\hline
			5 & $|0\rangle|0\rangle|0\rangle|1\rangle|0\rangle|1\rangle|1\rangle$ \\
			\hline
			6 & $|0\rangle|0\rangle|0\rangle|1\rangle|1\rangle|1\rangle|0\rangle$ \\
			\hline
			7 & $|0\rangle|0\rangle|0\rangle|1\rangle|1\rangle|1\rangle|1\rangle$ \\
			\hline
			8 & $|0\rangle|0\rangle|1\rangle|0\rangle|0\rangle|0\rangle|0\rangle$ \\
			\hline
			9 & $|0\rangle|1\rangle|1\rangle|0\rangle|0\rangle|0\rangle|0\rangle$ \\
			\hline
			10 & $|1\rangle|0\rangle|1\rangle|0\rangle|0\rangle|0\rangle|0\rangle$ \\
			\hline
			11 & $|1\rangle|1\rangle|1\rangle|0\rangle|0\rangle|0\rangle|0\rangle$ \\
			\hline
			12 & $|0\rangle|0\rangle|1\rangle|0\rangle|1\rangle|0\rangle|0\rangle$ \\
			\hline
			13 & $|1\rangle|0\rangle|1\rangle|0\rangle|1\rangle|0\rangle|0\rangle$ \\
			\hline
			14 & $|0\rangle|0\rangle|1\rangle|1\rangle|0\rangle|0\rangle|0\rangle$ \\
			\hline
			15 & $|0\rangle|1\rangle|1\rangle|1\rangle|0\rangle|0\rangle|0\rangle$ \\
			\hline
			16 & $|0\rangle|0\rangle|1\rangle|1\rangle|1\rangle|0\rangle|0\rangle$ \\
			\hline
		\end{tabular}
		\caption{{\it Quantum states involving in evolution.}}	
		\label{tab:States}
	\end{table}
	
	The photon annihilation and creation operators on an $p_1$-photon state with mode $\Omega^{\uparrow}$ are described as
	\begin{equation}
		\label{eq:PhotonOperatorsPhotonUp}
		\begin{aligned}
			&if\ p_1>0,\ \left \{
				\begin{aligned}
				&a_{\Omega^{\uparrow}}|p_1\rangle_{\Omega^{\uparrow}}=\sqrt{p_1}|p_1-1\rangle_{\Omega^{\uparrow}},\\
				&a_{\Omega^{\uparrow}}^{\dag}|p_1\rangle_{\Omega^{\uparrow}}=\sqrt{p_1+1}|p_1+1\rangle_{\Omega^{\uparrow}},
				\end{aligned}
				\right
				.\\
			&if\ p_1=0,\ \left \{
				\begin{aligned}
					&a_{\Omega^{\uparrow}}|0\rangle_{\Omega^{\uparrow}}=0,\\
					&a_{\Omega^{\uparrow}}^{\dag}|0\rangle_{\Omega^{\uparrow}}=|1\rangle_{\Omega^{\uparrow}}.
				\end{aligned}
				\right
				.\\
		\end{aligned}
	\end{equation}
	And their corresponding operators for electronic transitions, $\sigma_{\Omega^{\uparrow}},\ \sigma_{\Omega^{\uparrow}}^{\dag}$, have following form
	\begin{equation}
		\label{eq:ElecTransOperatorsUp}
		\begin{aligned}
			&\sigma_{\Omega^{\uparrow}}|0\rangle_{\Phi^{\uparrow}}=0,\\
			&\sigma_{\Omega^{\uparrow}}|1\rangle_{\Phi^{\uparrow}}=|0\rangle_{\Phi^{\uparrow}},\\
			&\sigma_{\Omega^{\uparrow}}^{\dag}|0\rangle_{\Phi^{\uparrow}}=|1\rangle_{\Phi^{\uparrow}},\\
			&\sigma_{\Omega^{\uparrow}}^{\dag}|1\rangle_{\Phi^{\uparrow}}=0.
		\end{aligned}
	\end{equation}
	
	The photon annihilation and creation operators on an $p_2$-photon state with mode $\Omega^{\downarrow}$ are described as
	\begin{equation}
		\label{eq:PhotonOperatorsPhotonDown}
		\begin{aligned}
			&if\ p_2>0,\ \left \{
				\begin{aligned}
				&a_{\Omega^{\downarrow}}|p_2\rangle_{\Omega^{\downarrow}}=\sqrt{p_2}|p_2-1\rangle_{\Omega^{\downarrow}},\\
				&a_{\Omega^{\downarrow}}^{\dag}|p_2\rangle_{\Omega^{\downarrow}}=\sqrt{p_2+1}|p_2+1\rangle_{\Omega^{\downarrow}},
				\end{aligned}
				\right
				.\\
			&if\ p_2=0,\ \left \{
				\begin{aligned}
					&a_{\Omega^{\downarrow}}|0\rangle_{\Omega^{\downarrow}}=0,\\
					&a_{\Omega^{\downarrow}}^{\dag}|0\rangle_{\Omega^{\downarrow}}=|1\rangle_{\Omega^{\downarrow}}.
				\end{aligned}
				\right
				.\\
		\end{aligned}
	\end{equation}
	And their corresponding operators for electronic transitions, $\sigma_{\Omega^{\downarrow}},\ \sigma_{\Omega^{\downarrow}}^{\dag}$, have following form
	\begin{equation}
		\label{eq:ElecTransOperatorsDown}
		\begin{aligned}
			&\sigma_{\Omega^{\downarrow}}|0\rangle_{\Phi^{\downarrow}}=0,\\
			&\sigma_{\Omega^{\downarrow}}|1\rangle_{\Phi^{\downarrow}}=|0\rangle_{\Phi^{\downarrow}},\\
			&\sigma_{\Omega^{\downarrow}}^{\dag}|0\rangle_{\Phi^{\downarrow}}=|1\rangle_{\Phi^{\downarrow}},\\
			&\sigma_{\Omega^{\downarrow}}^{\dag}|1\rangle_{\Phi^{\downarrow}}=0.
		\end{aligned}
	\end{equation}
	
	The phonon annihilation and creation operators on an $m$-phonon state with mode $\omega$ are described as
	\begin{equation}
		\label{eq:PhononOperatorsPhonon}
		\begin{aligned}
			&if\ m>0,\ \left \{
				\begin{aligned}
				&a_{\omega}|m\rangle_{\omega}=\sqrt{m}|m-1\rangle_{\omega},\\
				&a_{\omega}^{\dag}|m\rangle_{\omega}=\sqrt{m+1}|m+1\rangle_{\omega},
				\end{aligned}
				\right
				.\\
			&if\ m=0,\ \left \{
				\begin{aligned}
					&a_{\omega}|0\rangle_{\omega}=0,\\
					&a_{\omega}^{\dag}|0\rangle_{\omega}=|1\rangle_{\omega}.
				\end{aligned}
				\right
				.\\
		\end{aligned}
	\end{equation}
	And the operators for formation-breaking of covalent bond, $\sigma_{\omega},\ \sigma_{\omega}^{\dag}$, have following form
	\begin{equation}
		\label{eq:CovalentBondOperators}
		\begin{aligned}
			&\sigma_{\omega}|0\rangle_{cb}=0,\\
			&\sigma_{\omega}|1\rangle_{cb}=|0\rangle_{cb},\\
			&\sigma_{\omega}^{\dag}|0\rangle_{cb}=|1\rangle_{cb}\\
			&\sigma_{\omega}^{\dag}|1\rangle_{cb}=0.
		\end{aligned}
	\end{equation}
	
	The operators for tunneling process of nuclei, $\sigma_n,\ \sigma_n^{\dag}$, have following form
	\begin{equation}
		\label{eq:TunnelingOperators}
		\begin{aligned}
			&\sigma_n|0\rangle_n=0,\\
			&\sigma_n|1\rangle_n=|0\rangle_n,\\
			&\sigma_n^{\dag}|0\rangle_n=|1\rangle_n\\
			&\sigma_n^{\dag}|1\rangle_n=0.
		\end{aligned}
	\end{equation}
	
	Initial state $|\Psi_{initial}\rangle$ is shown in Fig. \ref{fig:Model} {\bf (c)}, which can be decomposed into the sum of four states
	\begin{equation}
		\label{eq:InitialDecomposed}
		|\Psi_{initial}\rangle=\frac{1}{2}\left(|\Phi_0^{\uparrow}\Phi_0^{\downarrow}\rangle+|\Phi_1^{\uparrow}\Phi_0^{\downarrow}\rangle-|\Phi_0^{\uparrow}\Phi_1^{\downarrow}\rangle-|\Phi_1^{\uparrow}\Phi_1^{\downarrow}\rangle\right)
	\end{equation}
	where
	\begin{subequations}
		\label{eq:PhiPhi}
		\begin{align}
			|\Phi_0^{\uparrow}\Phi_0^{\downarrow}\rangle=|0\rangle_{\Omega^{\uparrow}}|0\rangle_{\Omega^{\downarrow}}|0\rangle_{\omega}|0\rangle_{\Phi_1^{\uparrow}}|0\rangle_{\Phi_1^{\downarrow}}|1\rangle_{cb}|1\rangle_{n}\label{eq:Phi0Phi0}\\
			|\Phi_1^{\uparrow}\Phi_0^{\downarrow}\rangle=|0\rangle_{\Omega^{\uparrow}}|0\rangle_{\Omega^{\downarrow}}|0\rangle_{\omega}|1\rangle_{\Phi_1^{\uparrow}}|0\rangle_{\Phi_1^{\downarrow}}|1\rangle_{cb}|1\rangle_{n}\label{eq:Phi1Phi0}\\
			|\Phi_0^{\uparrow}\Phi_1^{\downarrow}\rangle=|0\rangle_{\Omega^{\uparrow}}|0\rangle_{\Omega^{\downarrow}}|0\rangle_{\omega}|0\rangle_{\Phi_1^{\uparrow}}|1\rangle_{\Phi_1^{\downarrow}}|1\rangle_{cb}|1\rangle_{n}\label{eq:Phi0Phi1}\\
			|\Phi_1^{\uparrow}\Phi_1^{\downarrow}\rangle=|0\rangle_{\Omega^{\uparrow}}|0\rangle_{\Omega^{\downarrow}}|0\rangle_{\omega}|1\rangle_{\Phi_1^{\uparrow}}|1\rangle_{\Phi_1^{\downarrow}}|1\rangle_{cb}|1\rangle_{n}\label{eq:Phi1Phi1}
		\end{align}
	\end{subequations}
		
	\section{Numerical method}
	\label{sec:Method}
	
	\subsection{Precise time step integration method}
	\label{subsec:PTSIM}
	
	\begin{figure*}
		\begin{center}
		\includegraphics[width=1.\textwidth]{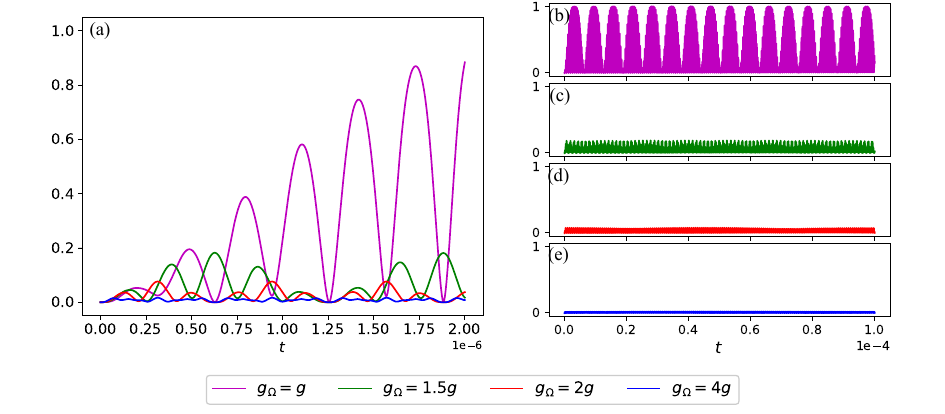}
		\end{center}
		\caption{(online color) {\it Effect of different strengths of electronic transition $g_{\Omega}$ on entropy.} The magenta, green, red and blue curves, respectively, stand for different electronic transition strengths.}
		\label{fig:ElecTrans}
	\end{figure*}
	
	We use the density matrix $\rho$ to obtain the unitary evolution of the non-dissipative system. The solution $\rho\left(t\right)$ may be approximately found as follows
	\begin{equation}
		\label{eq:UnitaryPart}
		\rho\left(t+dt\right)=exp\left({-\frac{i}{\hbar}Hdt}\right)\rho\left(t\right)exp\left(\frac{i}{\hbar}Hdt\right)
	\end{equation}
	where $H$ is defined in Eq. \eqref{eq:HamilBondPhonon}, and
	\begin{equation}
		\label{eq:InitialRho}
		\rho(t=0)=|\Psi_{initial}\rangle\langle\Psi_{initial}|
	\end{equation}
	where $|\Psi_{initial}\rangle$ is defined in Eq. \eqref{eq:InitialDecomposed}.
	
	In addition to being crucial for solving differential equations, the matrix exponential is also important for control theory, transportation and waveguides. The articles \cite{Moler1978, Moler2003} has examined 19 different ways to compute the matrix exponential and their drawbacks, and the fact that the exponential problem has not yet been fully resolved illustrates how important it is. Zhong Wanxie proposed the precise time step integration method (PTSIM) of matrix exponential in 1991 \cite{Zhong1991}. This method avoids computer truncation errors caused by fine division and improves the numerical solution of matrix exponential to computer accuracy. This method was quickly applied to solving mathematical and physical problems \cite{ZhongWilliams1994, Zhong1994, Zhong1995, GAO2016}. When solving the equations of motion using stepwise integration, the matrix exponential operation involved is $e^{A\Delta t}$. According to the addition theorem of matrix exponential
	\begin{equation}
		\label{eq:MatrixExp}
		e^{A\Delta t}=\left[e^{A\frac{\Delta t}{2^M}}\right]^{2^M}=\left[e^{A\varepsilon}\right]^{2^M}
	\end{equation}
	where $A$ is matrix, $\Delta t$ is time step, $\varepsilon=\frac{\Delta t}{2^M}$ (usually $M$ is taken to be equal to 20). The matrix exponential can be approximated using Taylor series expansion (4 terms)
	\begin{equation}
		\label{eq:Taylor}
		e^{A\varepsilon}\approx I+A\varepsilon+\frac{(A\varepsilon)^2}{2!}+\frac{(A\varepsilon)^3}{3!}+\frac{(A\varepsilon)^4}{4!}
	\end{equation}
	where $I$ is unit matrix, then
	\begin{equation}
		\label{eq:PTSIM}
		\begin{aligned}
			e^{A\Delta t}&\approx \left[I+A\varepsilon+\frac{(A\varepsilon)^2}{2!}+\frac{(A\varepsilon)^3}{3!}+\frac{(A\varepsilon)^4}{4!}\right]^{2^M}\\
			&=\left[I+T_{a,0}\right]^{2^M}
		\end{aligned}
	\end{equation}
	where $T_{a,0}=A\varepsilon+\frac{(A\varepsilon)^2}{2!}+\frac{(A\varepsilon)^3}{3!}+\frac{(A\varepsilon)^4}{4!}$, then
	\begin{equation}
		\label{eq:Recursion}
		\begin{aligned}
			\left[I+T_{a,0}\right]^{2^M}&=\left[(I+T_{a,0})^2\right]^{2^{M-1}}\\
			&=\left[I+2T_{a,0}+T_{a,0}T_{a,0}\right]^{2^{M-1}}\\
			&=\left[I+T_{a,1}\right]^{2^{M-1}}\\
			&=\left[I+T_{a,2}\right]^{2^{M-2}}\\
			&\cdots\cdots\\
			&=\left[I+T_{a,M}\right]
		\end{aligned}
	\end{equation}
	where $T_{a,n}=2T_{a,n-1}+T_{a,n-1}T_{a,n-1},\ n\geq1$. The above calculation process avoids operations between values with large differences in magnitude and avoids the impact of rounding errors.
	
	\subsection{Quantum entropy}
	\label{subsec:Entropy}
	
	\begin{figure*}
		\begin{center}
		\includegraphics[width=1.\textwidth]{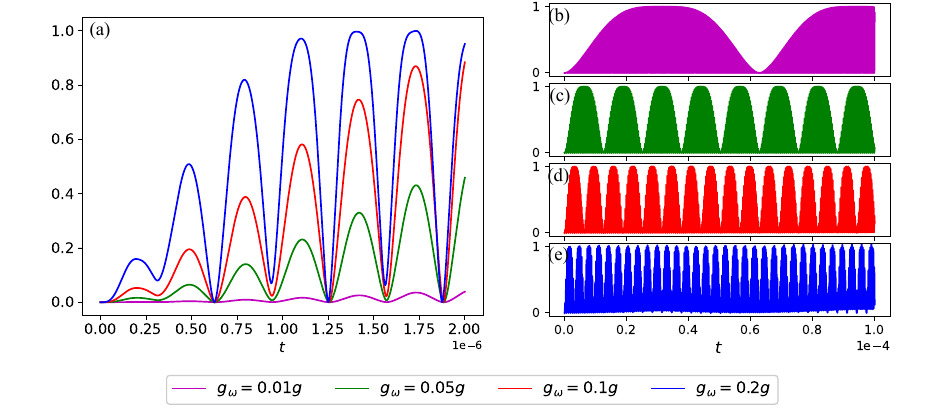}
		\end{center}
		\caption{(online color) {\it Effect of different strengths of covalent bond formation $g_{\omega}$ on entropy.} The magenta, green, red and blue curves, respectively, stand for different covalent bond formation strengths.}
		\label{fig:CovBond}
	\end{figure*}
	
	\begin{figure*}
		\begin{center}
		\includegraphics[width=1.\textwidth]{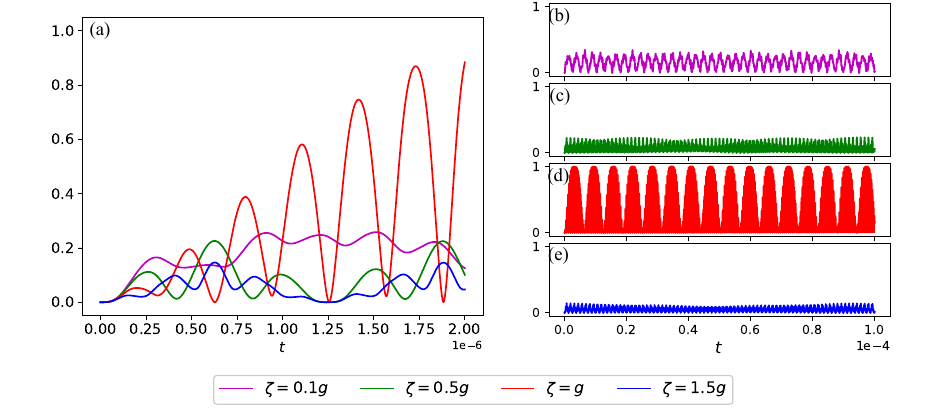}
		\end{center}
		\caption{(online color) {\it Effect of different tunneling effect strengths $\zeta$ on entropy.} The magenta, green, red and blue curves, respectively, stand for different tunneling effect strengths.}
		\label{fig:Tunneling}
	\end{figure*}
	
	For a bipartite system, we set the one subsystem as $\mathcal{A}$ and the another subsystem as $\mathcal{B}$. We exclusively address the entanglement in the non-dissipative situation in simulations. Quantum entropy was first introduced by von Neumann \cite{VonNeumann1932}. Its status is equivalent to that of Shannon entropy in classical information theory. It is the most basic and important concept in the fields of quantum information and quantum communication. From a thermodynamic point of view, the entropy of a closed system is a time-independent constant. This means that if a closed system is prepared in a pure (mixed) state, it will remain in a pure (mixed) state during the time evolution. But people are more interested in the interaction between a system and its surrounding environment, or the interaction between subsystems in a system. The density matrix describing the whole system cannot directly display the dynamics of each subsystem, so a statistical operator is needed to describe the subsystem. This is the reduced density operator of the subsystem, which can define the entropy of the subsystem and describe the dynamics of the subsystem. The von Neumann entropy of reduced density matrix, which is used to measure quantum correlation \cite{Obada2007, Obada2008}, can be used directly when the quantum system is closed and the entire system is in the pure state $|pure\rangle$. However, in this case, it can also be thought of as the entropy of entanglement. In this case, due to the Schmidt decomposition, the von Neumann entropy of each of two subsystems $\mathcal{A}$ and $\mathcal{B}$ can be used as a gauge of the system's entanglement \cite{Bennett1996}
	\begin{equation}
		\label{eq:vonNeumannEntropy}
		E=S(\mathcal{A})=S(\mathcal{B})
	\end{equation}
	where $S(\mathcal{A}),\ S(\mathcal{B})$ are corresponding to $\rho_{\mathcal{A}},\ \rho_{\mathcal{B}}$, respectively. And $\rho_{\mathcal{A}(\mathcal{B})}$ --- reduction of density matrix $\rho_{\mathcal{AB}}=|pure\rangle\langle pure|$, which is the density matrix of whole system, has the following form
	\begin{subequations}
		\label{eq:ReductionRho}
		\begin{align}
			\rho_{\mathcal{A}}&=Tr_{\mathcal{B}}(\rho_{\mathcal{AB}})\nonumber\\
			&=\sum_b(I_{\mathcal{A}}\otimes\langle b|_{\mathcal{B}})\rho_{\mathcal{AB}}(I_{\mathcal{A}}\otimes|b\rangle_{\mathcal{B}})\label{eq:ReductionRhoA}\\
			\rho_{\mathcal{B}}&=Tr_{\mathcal{A}}(\rho_{\mathcal{AB}})\nonumber\\
			&=\sum_a(\langle a|_{\mathcal{A}}\otimes I_{\mathcal{B}})\rho_{\mathcal{AB}}(|a\rangle_{\mathcal{A}}\otimes I_{\mathcal{B}})\label{eq:ReductionRhoB}
		\end{align}
	\end{subequations}
	where $Tr_{\mathcal{A},\mathcal{B}}$ represents the trace operation on the variables of the subsystem. Some details for calculating reduced density matrix and its matrix form are shown in Appx. \ref{appx:Reduced}.
	
	In quantum theory, the definition of von Neumann entropy of reduced density matrix $\rho_{\mathcal{A}(\mathcal{B})}$, called quantum reduced entropy, is as follows
	\begin{equation}
		\label{eq:vonNeumannTrace}
		S(\mathcal{A}(\mathcal{B}))=-Tr(\rho_{\mathcal{A}(\mathcal{B})} log_2\rho_{\mathcal{A}(\mathcal{B})})
	\end{equation}
	 In terms of eigenvalues, we have
	\begin{equation}
		\label{eq:vonNeumannEigen}
		S(\mathcal{A}(\mathcal{B}))=-\sum_i\lambda_i^{\mathcal{A}(\mathcal{B})}log_2\lambda_i^{\mathcal{A}(\mathcal{B})}
	\end{equation}
	where $\lambda_i^{\mathcal{A}(\mathcal{B})}$ --- eigenvalues of density matrix $\rho_{\mathcal{A}(\mathcal{B})}$. $0\leq S(\mathcal{A}(\mathcal{B}))\leq log_2N,\ N=2^n,\ N$ --- Hilbert space dimension, $n$ --- number of qubits of subsystem. $S(\mathcal{A}(\mathcal{B}))=0$ --- separable state, $S(\mathcal{A}(\mathcal{B}))>0$ --- entangled state, $S(\mathcal{A}(\mathcal{B}))=log_2N$ --- maximum entangled state.

	\section{Simulations and results} 
	\label{sec:Simulation}
	
	\begin{figure}
		\begin{center}
		\includegraphics[width=0.5\textwidth]{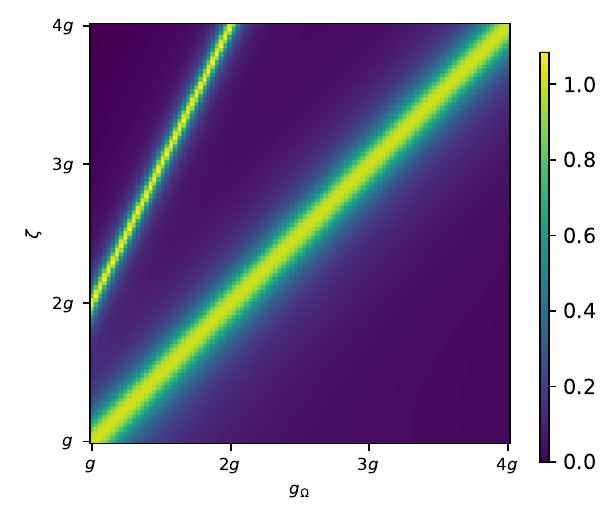}
		\end{center}
		\caption{(online color) {\it Effect of different $g_{\Omega}$ and $\zeta$ on entropy.} Each point represents the maximum of entropy of unitary evolution under the corresponding $g_{\Omega}$ and $\zeta$.}
		\label{fig:ElecTransTunneling}
	\end{figure}
	
	\begin{figure}
		\begin{center}
		\includegraphics[width=0.5\textwidth]{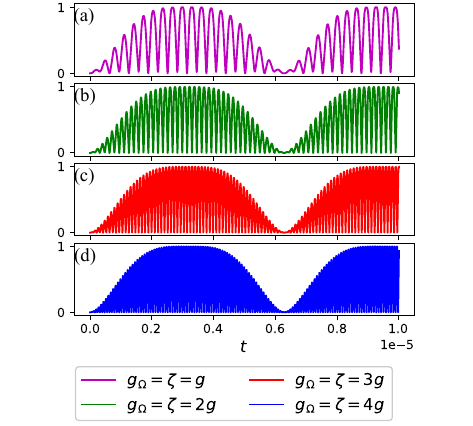}
		\end{center}
		\caption{(online color) {\it Effect of different $g_{\Omega}$ and $\zeta$ on entropy when $\zeta=g_{\Omega}$.} (a) $\zeta=g_{\Omega}=g$ (magenta); (b) $\zeta=g_{\Omega}=2g$ (green); (c) $\zeta=g_{\Omega}=3g$ (red); (d) $\zeta=g_{\Omega}=4g$ (blue).}
		\label{fig:ElecTransTunneling1}
	\end{figure}

	\begin{figure}
		\begin{center}
		\includegraphics[width=0.5\textwidth]{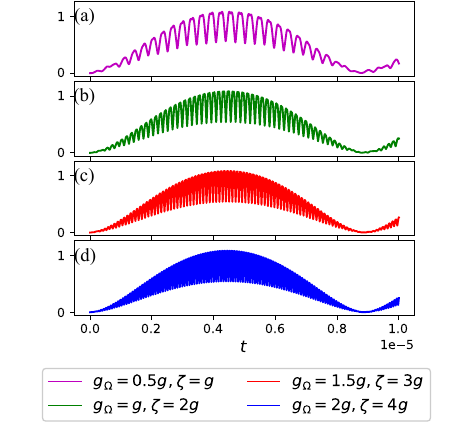}
		\end{center}
		\caption{(online color) {\it Effect of different $g_{\Omega}$ and $\zeta$ on entropy when $\zeta=2g_{\Omega}$.} (a) $\zeta=g,\ g_{\Omega}=0.5g$ (magenta), (b) $\zeta=2g,\ g_{\Omega}=g$ (green), (c) $\zeta=3g,\ g_{\Omega}=1.5g$ (red), (d) $\zeta=4g,\ g_{\Omega}=2g$ (blue).}
		\label{fig:ElecTransTunneling2}
	\end{figure}
	
	Our simulations consist of two parts: the effect of the interactions on the entropy shown in Sec. \ref{subsec:Effect} and the comparison of the entropy of different subsystems shown in Sec. \ref{subsec:Comparison}. In this paper we consider and fix following conditions in all simulations: $\hbar=1,\ \Omega^{\uparrow}=\Omega^{\downarrow}=10^9,\ \omega=10^8,\ g=10^7$.
	
	\subsection{Effect of interactions on entropy} 
	\label{subsec:Effect}
	
	We divide the whole system into the following two subsystems
		\begin{equation}
		\label{eq:Subsystems1}
		|\Psi\rangle_{\mathcal{C}}=\underbrace{|p_1\rangle_{\Omega^{\uparrow}}|p_2\rangle_{\Omega^{\downarrow}}}_{\mathcal{A}}\underbrace{|m\rangle_{\omega}|l_1\rangle_{\Phi_1^{\uparrow}}|l_2\rangle_{\Phi_1^{\downarrow}}|L\rangle_{cb}|k\rangle_{n}}_{\mathcal{B}}
	\end{equation}
	where $\mathcal{A}$ is a subsystem composed of photons, $\mathcal{B}$ is a system composed of matter. We define $S_{\Omega}=S(\mathcal{A})=S(\mathcal{B})$. We will study the effect of different values of $g_{\Omega},\ g_{\omega}, \zeta$ on entropy $S_{\Omega}$.
	
	Firstly, we focus on the effect of electronic transition $g_{\Omega}$ on entropy. We suppose $g_{\omega}=0.1g,\ \zeta=g$. Four different values of $g_{\Omega}$ are $g$, $1.5g$, $2g$ and $4g$. The time-dependent curve of entropy takes on a wave packet-like shape. In Fig. \ref{fig:ElecTrans} (a), we find that the greater the strength, the lower the peak value of entropy. When $g_{\Omega}$ equal to $g$, the peak value is approximately equal to $1$, and when $g_{\Omega}$ is equal to $4g$, the peak value is close to $0$. And in the four cases, the wave packets of entropy are not synchronized. In Fig. \ref{fig:ElecTrans} (b$\sim$e), we study the envelope of wave packet. We find that the greater the strength, the lower the maximum value of the envelope of the wave packet and the shorter its period.
	
	Secondly, we focus on the effect of covalent bond formation $g_{\omega}$ on entropy. We suppose $g_{\Omega}=g,\ \zeta=g$. Four different values of $g_{\omega}$ are $0.01g$, $0.05g$, $0.1g$ and $0.2g$. In Fig. \ref{fig:CovBond} (a), we find that different from Fig. \ref{fig:ElecTrans} (a), the oscillation frequencies of the entropy wave packets in these four cases are synchronized. In Fig. \ref{fig:CovBond} (b$\sim$e), we find that peak value in these four cases are almost the same and equal to $1$. However, the envelopes of wave packet are different. Obviously, the length of the period of envelope is strictly inversely proportional to the size of value of $g_{\omega}$.
	
	Thirdly, we focus on the effect of tunneling effect $\zeta$ on entropy. We suppose $g_{\Omega}=g,\ g_{\omega}=0.1g$. Four different values of $\zeta$ are $0.1g$, $0.5g$, $g$ and $1.5g$. In Fig. \ref{fig:Tunneling} (a), in the four cases, the wave packets of entropy are not synchronized as same as Fig. \ref{fig:ElecTrans} (a). Moreover, when $\zeta=g$, the peak value of entropy is maximum. In Fig. \ref{fig:Tunneling} (b$\sim$e), we can more obviously find that the peak value of entropy changes with the change of value of $\zeta$, and when $\zeta=g$, the peak value reaches a maximum of $1$. In addition, the length of the period of envelope also changes with the change of value of $\zeta$: the period reaches its maximum value when $\zeta=g$.
	
\begin{figure}
		\begin{center}
		\includegraphics[width=0.5\textwidth]{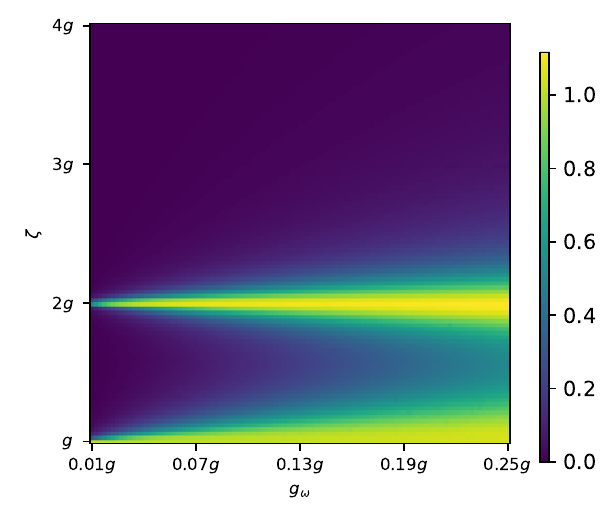}
		\end{center}
		\caption{(online color) {\it Effect of different $g_{\omega}$ and $\zeta$ on entropy.} Each point represents the maximum of entropy of unitary evolution under the corresponding $g_{\omega}$ and $\zeta$.}
		\label{fig:CovBondTunneling}
	\end{figure}

\begin{figure}
		\begin{center}
		\includegraphics[width=0.5\textwidth]{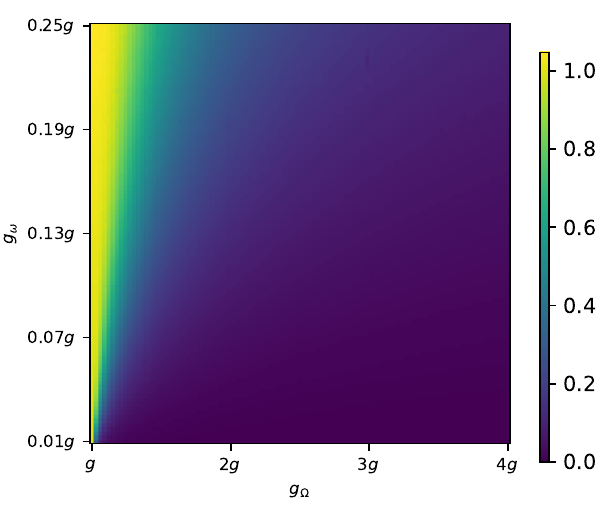}
		\end{center}
		\caption{(online color) {\it Effect of different $g_{\Omega}$ and $g_{\omega}$ on entropy when $\zeta=g$.} Each point represents the maximum of entropy of unitary evolution under the corresponding $g_{\Omega}$ and $g_{\omega}$.}
		\label{fig:ElecTransCovBond1G}
	\end{figure}

\begin{figure}
		\begin{center}
		\includegraphics[width=0.5\textwidth]{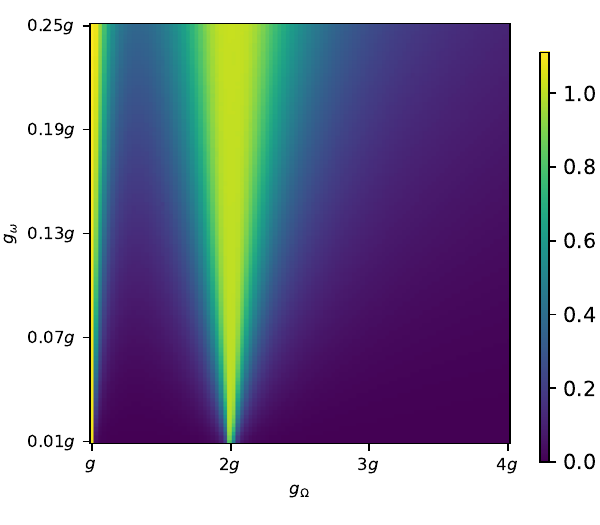}
		\end{center}
		\caption{(online color) {\it Effect of different $g_{\Omega}$ and $g_{\omega}$ on entropy  when $\zeta=2g$.} Each point represents the maximum of entropy of unitary evolution under the corresponding $g_{\Omega}$ and $g_{\omega}$.}
		\label{fig:ElecTransCovBond2G}
	\end{figure}

	Now we find that all electron transition strength, covalent bond formation strength and tunneling effect strength can affect the von Neumann entropy, so next we study the effect on entropy between electron transition and tunneling effect (see Fig. \ref{fig:ElecTransTunneling}), between covalent bond formation and tunneling effect (see Fig. \ref{fig:CovBondTunneling}), between electron transition and covalent bond formation with two different cases (see Figs. \ref{fig:ElecTransCovBond1G} and \ref{fig:ElecTransCovBond2G}). In these figures, each point represents the maximum of entropy of unitary evolution under the corresponding strengths.
	
	Firstly, we investigate the effect on entropy of different electron transition strength and tunneling effect strength. Here $g_{\omega}=0.1g$. In Fig. \ref{fig:ElecTransTunneling}, there are two obvious ridges on this figure. One is when $\zeta=g_{\Omega}$, the local maximum of points obtained on this ridge is equal to $1$, the another is when $\zeta=2g_{\Omega}$, the local maximum of points obtained on this ridge is greater than $1$. Interestingly, when we plot the time-evolution curves of the above two cases, we get two different wave packet types, shown in Figs. \ref{fig:ElecTransTunneling1} and \ref{fig:ElecTransTunneling2}, respectively. Comparing these two figures, we can find that the envelope period of the wave packet in Fig. \ref{fig:ElecTransTunneling1} is shorter than that in Fig. \ref{fig:ElecTransTunneling2}. In addition, when $\zeta=2g_{\Omega}$, the wave packet presents a bow shape: that is, the entropy will be equal to $0$ only at the initial stage and the end stage of the envelope period.
	
\begin{figure}
		\begin{center}
		\includegraphics[width=0.5\textwidth]{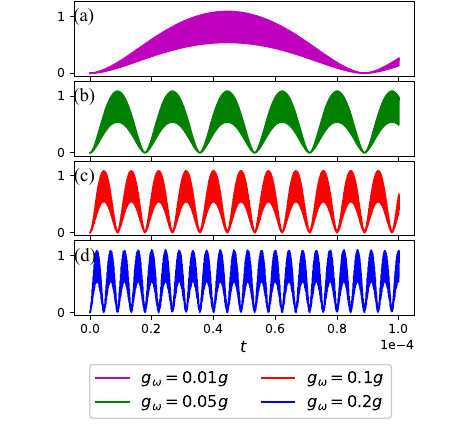}
		\end{center}
		\caption{(online color) {\it Effect of different $g_{\omega}$ and $\zeta$ on entropy when $\zeta=2g$.} (a) $g_{\omega}=0.01g$ (magenta), (b) $g_{\omega}=0.05g$ (green), (c) $g_{\omega}=0.1g$ (red), (d) $g_{\omega}=0.2g$ (blue).}
		\label{fig:CovBondTunneling1}
	\end{figure}

	Secondly, we investigate the effect on entropy of different covalent bond formation strength and tunneling effect strength. Here $g_{\Omega}=g$. In Fig. \ref{fig:CovBondTunneling}, there are two emission-like beams from left to right, and the local maximum of points is concentrated in these two beams. As $g_{\omega}$ becomes larger, the divergence of these two beams increases, and two beams cross at the right end of figure. Similar to Fig. \ref{fig:ElecTransTunneling}, when $\zeta=g$ and $\zeta=2g$, two ridges are obtained respectively, and local maximum of points are obtained on these ridges. When $\zeta=g_{\Omega}=g$, the local maximum of points obtained on this ridge is equal to $1$, and when $\zeta=2g_{\Omega}=2g$, the local maximum of points obtained on the ridge is greater than $1$. For the first case, we plot the time-evolution curves in Fig. \ref{fig:CovBond}, and now we plot the time-evolution curves of the second case in Fig. \ref{fig:CovBondTunneling1}. Comparing these two figures, we can find that the envelope period of the wave packet in Fig. \ref{fig:CovBond} is shorter than that in Fig. \ref{fig:CovBondTunneling1}, too. Similarly, the length of the period of envelope in Fig. \ref{fig:CovBondTunneling1} is strictly inversely proportional to the size of value of $g_{\omega}$, and when $\zeta=2g_{\Omega}=2g$, the wave packet also presents a bow shape.
	
	Thirdly, we investigate the effect on entropy of different electron transition strength and covalent bond formation strength. Considering that $\zeta=g$ and $\zeta=2g$ have different effects on entropic dynamics above, we study the effects of different $g_{\Omega}$ and $g_{\omega}$ on entropy at the same time in these two cases. Thus, when $\zeta=g$, Fig. \ref{fig:ElecTransCovBond1G} is obtained. We can find that there is an emission-like beam from bottom to top at the left end. When $\zeta=g_{\Omega}=g$, local maximum of points are obtained. In Fig. \ref{fig:ElecTransCovBond2G}, when $\zeta=2g$, we can find that there are two emission-like beams from bottom to top at both the left and right ends. When $\zeta=2g_{\Omega}=2g$, local maximum of points are obtained at the left end, and when $\zeta=g_{\Omega}=2g$, local maximum of points are obtained at the right of the first emission-like beam. Different from the two clear ridges in Fig. \ref{fig:ElecTransTunneling}, the reason for the emission-like beam shape in Figs. \ref{fig:ElecTransCovBond1G} and \ref{fig:ElecTransCovBond2G} is the increase in $g_{\omega}$.
	
	\subsection{Comparison of entropy of different subsystems} 
	\label{subsec:Comparison}
	
	\begin{figure*}
		\begin{center}
		\includegraphics[width=1.\textwidth]{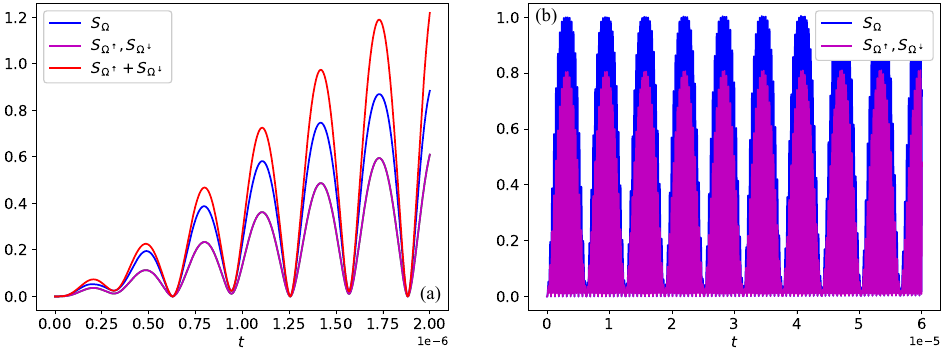}
		\end{center}
		\caption{(online color) {\it Comparison of von Neumann entropy of different photonic states.} $S_{\Omega^{\uparrow}}$ is entropy of photon with mode $\Omega^{\uparrow}$, $S_{\Omega^{\downarrow}}$ is entropy of photon with mode $\Omega^{\downarrow}$, $S_{\Omega}$ is entropy of both two types of photon.}
		\label{fig:ComparisonElecTrans}
	\end{figure*}
	
	\begin{figure*}
		\begin{center}
		\includegraphics[width=1.\textwidth]{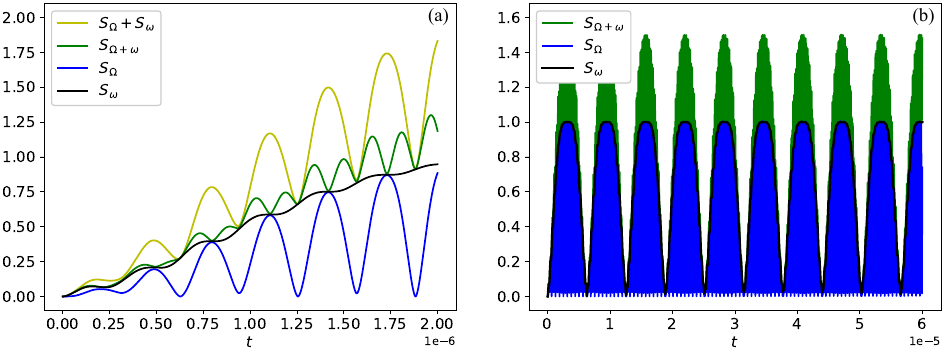}
		\end{center}
		\caption{(online color) {\it Comparison of von Neumann entropy between photonic state and phononic state.} $S_{\omega}$ is entropy of phonon with mode $\omega$.}
		\label{fig:ComparisonElecTransCovBond}
	\end{figure*}
	
	We assume the following situations in Fig. \ref{fig:ComparisonElecTrans}
	\begin{equation}
		\label{eq:Subsystems2}
		|\Psi\rangle_{\mathcal{C}}=\underbrace{|p_1\rangle_{\Omega^{\uparrow}}}_{\mathcal{A}}\underbrace{|p_2\rangle_{\Omega^{\downarrow}}|m\rangle_{\omega}|l_1\rangle_{\Phi_1^{\uparrow}}|l_2\rangle_{\Phi_1^{\downarrow}}|L\rangle_{cb}|k\rangle_{n}}_{\mathcal{B}}
	\end{equation}
	where $\mathcal{A}$ is a subsystem composed of photon with mode $\Omega^{\uparrow}$, and
	\begin{equation}
		\label{eq:Subsystems3}
		|\Psi\rangle_{\mathcal{C}}=\underbrace{|p_1\rangle_{\Omega^{\uparrow}}}_{\mathcal{B}}\underbrace{|p_2\rangle_{\Omega^{\downarrow}}}_{\mathcal{A}}\underbrace{|m\rangle_{\omega}|l_1\rangle_{\Phi_1^{\uparrow}}|l_2\rangle_{\Phi_1^{\downarrow}}|L\rangle_{cb}|k\rangle_{n}}_{\mathcal{B}}
	\end{equation}
	where $\mathcal{A}$ is a subsystem composed of photon with mode $\Omega^{\downarrow}$. We define $S_{\Omega^{\uparrow}},\ S_{\Omega^{\downarrow}}$ are entropies corresponding to photon with mode $\Omega^{\uparrow}$ and photon with mode $\Omega^{\downarrow}$, respectively.
	
	In Fig. \ref{fig:ComparisonElecTrans}, we get the following relational expression
	\begin{equation}
		\label{eq:EntropyInequality1}
		S_{\Omega^{\uparrow}}+S_{\Omega^{\downarrow}}\geq S_{\Omega}\geq S_{\Omega^{\uparrow}}=S_{\Omega^{\downarrow}}
	\end{equation}		
	where $S_{\Omega^{\uparrow}}$ always coincides with $S_{\Omega^{\downarrow}}$ since our quantum system is symmetrical, and the result satisfies that the sum of the entropies of the two subsystems $S_{\Omega^{\uparrow}}+S_{\Omega^{\downarrow}}$ is greater than the entire entropy of these subsystems $S_{\Omega}$. In Fig. \ref{fig:ComparisonElecTrans} (b), we can find that the wave packet envelopes of $S_{\Omega},\ S_{\Omega^{\uparrow}},\ S_{\Omega^{\downarrow}}$ are all synchronized.
	
	We assume the following situations in Fig. \ref{fig:ComparisonElecTransCovBond}
	\begin{equation}
		\label{eq:Subsystems4}
		|\Psi\rangle_{\mathcal{C}}=\underbrace{|p_1\rangle_{\Omega^{\uparrow}}|p_2\rangle_{\Omega^{\downarrow}}}_{\mathcal{B}}\underbrace{|m\rangle_{\omega}}_{\mathcal{A}}\underbrace{|l_1\rangle_{\Phi_1^{\uparrow}}|l_2\rangle_{\Phi_1^{\downarrow}}|L\rangle_{cb}|k\rangle_{n}}_{\mathcal{B}}
	\end{equation}
	where $\mathcal{A}$ is a subsystem composed of phonon with mode $\omega$, and
	\begin{equation}
		\label{eq:Subsystems5}
		|\Psi\rangle_{\mathcal{C}}=\underbrace{|p_1\rangle_{\Omega^{\uparrow}}|p_2\rangle_{\Omega^{\downarrow}}|m\rangle_{\omega}}_{\mathcal{A}}\underbrace{|l_1\rangle_{\Phi_1^{\uparrow}}|l_2\rangle_{\Phi_1^{\downarrow}}|L\rangle_{cb}|k\rangle_{n}}_{\mathcal{B}}
	\end{equation}
	where $\mathcal{A}$ is a subsystem composed of photons and phonon. We define $S_{\omega}$ is entropy corresponding to phonon with mode $\omega$, $S_{\Omega+\omega}$ is entropy corresponding to photons and phonon.
	
	In Fig. \ref{fig:ComparisonElecTransCovBond}, we get the following relational expression
	\begin{equation}
		\label{eq:EntropyInequality2}
		S_{\Omega}+S_{\omega}\geq S_{\Omega+\omega}\geq S_{\omega}\geq S_{\Omega}
	\end{equation}		
	where the sum of the entropies of the two subsystems $S_{\Omega}+S_{\omega}$ is greater than the entire entropy of these subsystems $S_{\Omega+\omega}$. In particular, the wave packets of $S_{\omega}$, $S_{\Omega+\omega}$ and $S_{\Omega}+S_{\omega}$ are bow-shaped. This means that all time-dependent curves of the entropy of subsystems containing phonons have bow shape. In Fig. \ref{fig:ComparisonElecTransCovBond} (b), we can find that the envelopes of $S_{\Omega+\omega},\ S_{\Omega},\ S_{\omega}$ are all synchronized.
	
	\section{Concluding discussion and future work} 
	\label{sec:ConcluFuture}
	
	In this paper, we explore various aspects of entropy of a multiqubit cavity QED system. The entropic dynamics has been constructed, and several analytical findings have been drawn from it:
	
	In Sec. \ref{subsec:Effect}, according to results from Figs. \ref{fig:ElecTrans}, \ref{fig:CovBond} and \ref{fig:Tunneling}, we proved that $g_{\Omega}$ and $\zeta$ affect significantly the peak value of entropy, but $g_{\omega}$ affects slightly that. Different values of $g_{\omega}$ will not affect the synchronization of wave packets, but changes in the values of $g_{\Omega}$ or $\zeta$ will affect the synchronization. According to results from Figs. \ref{fig:ElecTransTunneling}, \ref{fig:CovBondTunneling}, \ref{fig:ElecTransCovBond1G} and \ref{fig:ElecTransCovBond2G}, we will find that when $\zeta$ is equal to or twice as large as $g_{\Omega}$, the local maximum of points can be obtained. In particular, when $\zeta$ is twice as large as $g_{\Omega}$, the entropy curves take on a bow shape (see Figs. \ref{fig:ElecTransTunneling2} and \ref{fig:CovBondTunneling1}). In addition, the strength of the interaction will also change the period of the wave packet envelope, unless $g_{\Omega}$ and $\zeta$ are increased in equal proportions.
	
	In Sec. \ref{subsec:Comparison}, we compared the entropic dynamics between different subsystems. We obtained two inequalities between the entropies of different subsystems (see Eqs. \eqref{eq:EntropyInequality1}, \eqref{eq:EntropyInequality2}), and also verified that the sum of the entropies of the two subsystems is greater than or equal to the entropy of the entire system composed of these two subsystems. We find that the wave packet of the entropy of any subsystem containing phonon exhibits a bow shape.
	
	We tried to study the characteristics of entropy on a seven-qubit complicated quantum system and obtained some regular results. The results show that the entropic dynamics can be controlled by selectively choosing system parameters, and some rules are obtained. These results lay the foundation for our future studies of more complicated quantum model entropy dynamics.
	
	In fact, in the real physical system of two hydrogen atoms, two hydrogen atoms can be placed in an optical cavity --- the Fabry--Pérot resonator, by optical tweezers. The interaction between the hydrogen atoms and the two different fields in the cavity completes the transition of electrons and the formation and breaking of covalent bonds. We know that the transition of hydrogen atom electrons from the ground state to the first excited state requires the absorption of a photon with an energy of 10.2 eV, and the breaking of the covalent bond of the hydrogen molecule requires the absorption of a phonon with an energy of 4.5 eV. Since in the paper we assume that the electron transition energy is greater than the covalent bond energy of the hydrogen molecule. Similarly, in the real physical system, the electron transition energy is also greater than the covalent bond energy. Thus, we can predict the entanglement dynamics in this real physical system is similar to the results obtained in our paper.
	
	\begin{acknowledgments}
	This work was supported by the China Scholarship Council (CSC No.202108090483). The author acknowledges Center for Collective Usage of Ultra HPC Resources (https://www.parallel.ru/) at Lomonosov Moscow State University for providing supercomputer resources that have contributed to the research results reported within this paper.
	\end{acknowledgments}

\onecolumngrid

\appendix

	\section{Rotating wave approximation}
	\label{appx:RWA}
	
	RWA is taken into account in this paper. This approach ignores the quickly oscillating terms $\sigma^{\dag}a^{\dag},\ \sigma a$ in a Hamiltonian. When the strength of the applied electromagnetic radiation is close to resonance with an atomic transition and the strength is low, this approximation holds true \cite{Wu2007}. Thus,
	\begin{equation}
		\label{appxeq:RWACondition}
		\frac{g}{\hbar\omega_{cavity}}\approx\frac{g}{\hbar\omega_{atom}}\ll 1
	\end{equation}
where $\omega_{cavity}$ stands for cavity frequency, which includes $\Omega_c$ ($\Omega^{\uparrow}_c$ and $\Omega^{\downarrow}_c$) for photon and $\omega_c$ for phonon; and $\omega_{atom}$ for transition frequency (of atom), which includes $\Omega_a$ ($\Omega^{\uparrow}_a$ and $\Omega^{\downarrow}_a$) for electronic transition and $\omega_a$ formation-break of covalent bond. RWA allows us to change $\left(\sigma^{\dag}+\sigma\right)\left(a^{\dag}+a\right)$ to $\sigma^{\dag}a+\sigma a^{\dag}$. In Eq. \eqref{eq:HamilBondPhonon} we typically presume that $\Omega_{cavity}=\Omega_{atom}$: $\Omega_c=\Omega_a=\Omega$ ($\Omega^{\uparrow}_c=\Omega^{\uparrow}_a=\Omega^{\uparrow}$ and $\Omega^{\downarrow}_c=\Omega^{\downarrow}_a=\Omega^{\downarrow}$), $\omega_c=\omega_a=\omega$.

	\section{Matrix form of operators}
	\label{appx:Operators}
	
	The matrix form of annihilation and creation operators $a$ and $a^{\dag}$ in Eq. \eqref{eq:HamilBondPhonon} are described as follows
	\begin{subequations}
		\label{appxeq:OperatorsA}
		\begin{align}
			a&=\begin{array}{c@{\hspace{-5pt}}l}
			 \begin{array}{c}
			 	|0\rangle \\
			 	|1\rangle \\
			 	|2\rangle \\
			 	\vdots \\
			 	\vdots \\
			 	|p-2\rangle \\
			 	|p-1\rangle \\
			 	|p\rangle \\
			 \end{array}
			 & \left(
			 \begin{array}{cccccccc}
			 	0 & 1 & 0 & \cdots & \cdots & 0 & 0 & 0 \\
			 	0 & 0 & \sqrt{2} & \cdots & \cdots & 0 & 0 & 0 \\
			 	0 & 0 & 0 & \cdots & \cdots & 0 & 0 & 0 \\
			 	\vdots & \vdots & \vdots & \ddots & \ddots & \vdots & \vdots & \vdots \\
			 	\vdots & \vdots & \vdots & \ddots & \ddots & \vdots & \vdots & \vdots \\
			 	0 & 0 & 0 & \cdots & \cdots & 0 & \sqrt{p-1} & 0 \\
			 	0 & 0 & 0 & \cdots & \cdots & 0 & 0 & \sqrt{p}\\
			 	0 & 0 & 0 & \cdots & \cdots & 0 & 0 & 0 \\
			 \end{array}
			 \right)
			\end{array}\label{appxeq:Annihilation}\\
			a^{\dag}&=\begin{array}{c@{\hspace{-5pt}}l}
			 \begin{array}{c}
			 	|0\rangle \\
			 	|1\rangle \\
			 	|2\rangle \\
			 	\vdots \\
			 	\vdots \\
			 	|p-2\rangle \\
			 	|p-1\rangle \\
			 	|p\rangle \\
			 \end{array}
			 & \left(
			 \begin{array}{cccccccc}
			 	0 & 0 & 0 & \cdots & \cdots & 0 & 0 & 0 \\
			 	1 & 0 & 0 & \cdots & \cdots & 0 & 0 & 0 \\
			 	0 & \sqrt{2} & 0 & \cdots & \cdots & 0 & 0 & 0 \\
			 	\vdots & \vdots & \vdots & \ddots & \ddots & \vdots & \vdots & \vdots \\
			 	\vdots & \vdots & \vdots & \ddots & \ddots & \vdots & \vdots & \vdots \\
			 	0 & 0 & 0 & \cdots & \cdots & 0 & 0 & 0 \\
			 	0 & 0 & 0 & \cdots & \cdots & \sqrt{p-1} & 0 & 0 \\
			 	0 & 0 & 0 & \cdots & \cdots & 0 & \sqrt{p} & 0 \\
			 \end{array}
			 \right)
			\end{array}\label{appxeq:Creation}
		\end{align}
	\end{subequations}
	where $p$ can be replaced by $p_1$, $p_2$ and $m$ in this paper. And the matrix form of relaxation and excitation operators $\sigma$ and $\sigma^{\dag}$ in Eq. \eqref{eq:HamilBondPhonon} are described as follows
	\begin{subequations}
		\label{appxeq:OperatorsSigma}
		\begin{align}
			\sigma&=\begin{array}{c@{\hspace{-5pt}}l}
			 \begin{array}{c}
			 	|0\rangle \\
			 	|1\rangle \\
			 \end{array}
			 & \left(
			 \begin{array}{cc}
			 	0 & 1 \\
			 	0 & 0 \\
			 \end{array}
			 \right)
			\end{array}\label{appxeq:Relaxation}\\
			\sigma^{\dag}&=\begin{array}{c@{\hspace{-5pt}}l}
			 \begin{array}{c}
			 	|0\rangle \\
			 	|1\rangle \\
			 \end{array}
			 & \left(
			 \begin{array}{cc}
			 	0 & 0 \\
			 	1 & 0 \\
			 \end{array}
			 \right)
			\end{array}\label{appxeq:Excitation}
		\end{align}
	\end{subequations}
	In this paper, operators $\sigma_{\Omega^{\uparrow}}$, $\sigma_{\Omega^{\downarrow}}$, $\sigma_{\omega}$ and $\sigma_n$, and their conjugate operators all obey the rule of Eq. \eqref{appxeq:OperatorsSigma}.
	
	\section{Reduced density matrix}
	\label{appx:Reduced}
	
	For reduced density matrix, we usually have three types of it: left-end reduced, right-end reduced and intermediate reduced, which are corresponding to these three following cases
	\begin{subequations}
		\begin{align}
			&|\Psi\rangle_{\mathcal{C}}=\underbrace{|\cdots\rangle}_{\mathcal{A}}|\cdots\cdots\rangle\\
			&|\Psi\rangle_{\mathcal{C}}=|\cdots\cdots\rangle\underbrace{|\cdots\rangle}_{\mathcal{A}}\\
			&|\Psi\rangle_{\mathcal{C}}=|\cdots\rangle\underbrace{|\cdots\rangle}_{\mathcal{A}}|\cdots\rangle
		\end{align}
	\end{subequations}

	Let’s give an example of three-qubit system, having following Hilbert space
	\begin{equation}
		\label{appxeq:ThreeQubit}
		|\Psi\rangle_{\mathcal{C}}=|x\rangle_0|y\rangle_1|z\rangle_2
	\end{equation}
	where $x,\ y,\ z\in\{0, 1\}$. And the density matrix is as follows
	\begin{equation}
		\label{appxeq:Rho012}
		\rho_{012}=\begin{array}{c@{\hspace{-5pt}}l}
		 \begin{array}{c}
		 	|000\rangle \\
		 	|001\rangle \\
		 	|010\rangle \\
		 	|011\rangle \\
		 	|100\rangle \\
		 	|101\rangle \\
		 	|110\rangle \\
		 	|111\rangle \\
		 \end{array}
		 & \left(
		 \begin{array}{cccccccc}
		 	a_{00} & a_{01} & a_{02} & a_{03} & a_{04} & a_{05} & a_{06} & a_{07} \\
		 	a_{10} & a_{11} & a_{12} & a_{13} & a_{14} & a_{15} & a_{16} & a_{17} \\
		 	a_{20} & a_{21} & a_{22} & a_{23} & a_{24} & a_{25} & a_{26} & a_{27} \\
		 	a_{30} & a_{31} & a_{32} & a_{33} & a_{34} & a_{35} & a_{36} & a_{37} \\
		 	a_{40} & a_{41} & a_{42} & a_{43} & a_{44} & a_{45} & a_{46} & a_{47} \\
		 	a_{50} & a_{51} & a_{52} & a_{53} & a_{54} & a_{55} & a_{56} & a_{57} \\
		 	a_{60} & a_{61} & a_{62} & a_{63} & a_{64} & a_{65} & a_{66} & a_{67} \\
		 	a_{70} & a_{71} & a_{72} & a_{73} & a_{74} & a_{75} & a_{76} & a_{77} \\
		 \end{array}
		 \right)
		\end{array}
	\end{equation}
	
	We will separately find the reduced density matrices $\rho_0$, $\rho_1$ and $\rho_2$. Firstly, according to Eq. \eqref{eq:ReductionRhoA}, reduced density matrix $\rho_{01}$ has following matrix form
	\begin{equation}
		\label{appxeq:Rho01}
		\rho_{01}=\begin{array}{c@{\hspace{-5pt}}l}
		 \begin{array}{c}
		 	|00\rangle \\
		 	|01\rangle \\
		 	|10\rangle \\
		 	|11\rangle \\
		 \end{array}
		 & \left(
		 \begin{array}{cccc}
		 	a_{00}+a_{11} & a_{02}+a_{13} & a_{04}+a_{15} & a_{06}+a_{17} \\
		 	a_{20}+a_{31} & a_{22}+a_{33} & a_{24}+a_{35} & a_{26}+a_{37} \\
		 	a_{40}+a_{51} & a_{42}+a_{53} & a_{44}+a_{55} & a_{46}+a_{57} \\
		 	a_{60}+a_{71} & a_{62}+a_{73} & a_{64}+a_{75} & a_{66}+a_{77} \\
		 \end{array}
		 \right)
		\end{array}
	\end{equation}
	Immediately followed, $\rho_{0}$ has following matrix form
	\begin{equation}
		\label{appxeq:Rho0}
		\rho_{0}=\begin{array}{c@{\hspace{-5pt}}l}
		 \begin{array}{c}
		 	|0\rangle \\
		 	|1\rangle \\
		 \end{array}
		 & \left(
		 \begin{array}{cc}
		 	a_{00}+a_{11}+a_{22}+a_{33} & a_{04}+a_{15}+a_{26}+a_{37} \\
		 	a_{40}+a_{51}+a_{62}+a_{73} & a_{44}+a_{55}+a_{66}+a_{77} \\
		 \end{array}
		 \right)
		\end{array}
	\end{equation}
	Then, according to Eq. \eqref{eq:ReductionRhoB}, reduced density matrix $\rho_{12}$ has following matrix form
	\begin{equation}
		\label{appxeq:Rho12}
		\rho_{12}=\begin{array}{c@{\hspace{-5pt}}l}
		 \begin{array}{c}
		 	|00\rangle \\
		 	|01\rangle \\
		 	|10\rangle \\
		 	|11\rangle \\
		 \end{array}
		 & \left(
		 \begin{array}{cccc}
		 	a_{00}+a_{44} & a_{01}+a_{45} & a_{02}+a_{46} & a_{03}+a_{47} \\
		 	a_{10}+a_{54} & a_{11}+a_{55} & a_{12}+a_{56} & a_{13}+a_{57} \\
		 	a_{20}+a_{64} & a_{21}+a_{65} & a_{22}+a_{66} & a_{23}+a_{67} \\
		 	a_{30}+a_{74} & a_{31}+a_{75} & a_{32}+a_{76} & a_{33}+a_{77} \\
		 \end{array}
		 \right)
		\end{array}
	\end{equation}
	Immediately followed, $\rho_{2}$ has following matrix form
	\begin{equation}
		\label{appxeq:Rho2}
		\rho_{2}=\begin{array}{c@{\hspace{-5pt}}l}
		 \begin{array}{c}
		 	|0\rangle \\
		 	|1\rangle \\
		 \end{array}
		 & \left(
		 \begin{array}{cc}
		 	a_{00}+a_{44}+a_{22}+a_{66} & a_{01}+a_{45}+a_{23}+a_{67} \\
		 	a_{10}+a_{54}+a_{32}+a_{76} & a_{11}+a_{55}+a_{33}+a_{77} \\
		 \end{array}
		 \right)
		\end{array}
	\end{equation}
	In Eq. \eqref{appxeq:Rho12}, $\rho_{12}$ is obtained by applying Eq. \eqref{eq:ReductionRhoB}. Next, by applying Eq. \eqref{eq:ReductionRhoA}, reduced density matrix $\rho_{1}$ is reduced from $\rho_{12}$
	\begin{equation}
		\label{appxeq:Rho1}
		\rho_{1}=\begin{array}{c@{\hspace{-5pt}}l}
		 \begin{array}{c}
		 	|0\rangle \\
		 	|1\rangle \\
		 \end{array}
		 & \left(
		 \begin{array}{cc}
		 	a_{00}+a_{44}+a_{11}+a_{55} & a_{02}+a_{46}+a_{13}+a_{57} \\
		 	a_{20}+a_{64}+a_{31}+a_{75} & a_{22}+a_{66}+a_{33}+a_{77} \\
		 \end{array}
		 \right)
		\end{array}
	\end{equation}
	For obtaining $\rho_1$, we can use another method: swapping the qubits' positions. For example, the Eq. \eqref{appxeq:ThreeQubit} can be rewritten as
	\begin{equation}
		\label{appxeq:ThreeQubitNew}
		|\Psi\rangle_{\mathcal{C}'}=|y\rangle_1|x\rangle_0|z\rangle_2
	\end{equation}
	where $\mathcal{C}\neq\mathcal{C}'$, and the new density matrix is $\rho_{102}$. Thus, we just apply twice the Eq. \eqref{eq:ReductionRhoA}, and $\rho_1$ is obtained. Here $\rho_{01}$ and $\rho_0$ are left-end reduced density matrices, $\rho_{12}$ and $\rho_2$ --- right-end reduced, $\rho_1$ --- intermediate reduced.
	
	Process of obtaining these three types of reduced density matrix has obvious regularities. We can extend them to the seven-qubit system, then each reduced density matrix $\mathcal{A}$ from Eqs. \eqref{eq:Subsystems1}, \eqref{eq:Subsystems2}, \eqref{eq:Subsystems3}, \eqref{eq:Subsystems4} and \eqref{eq:Subsystems5} can be obtained.

\twocolumngrid

\bibliography{bibliography}
	
\end{document}